\documentclass[conference]{IEEEtran}
\IEEEoverridecommandlockouts

\usepackage{cite}
\usepackage{amsmath,amssymb,amsfonts}
\usepackage{algorithmic}
\usepackage{graphicx}
\usepackage{textcomp}
\usepackage{xcolor}
\usepackage{subcaption}
\usepackage{mathtools}
\usepackage{multirow}

\definecolor{tueScharlaken}{HTML}{C81919}

\def\BibTeX{{\rm B\kern-.05em{\sc i\kern-.025em b}\kern-.08em
    T\kern-.1667em\lower.7ex\hbox{E}\kern-.125emX}}
\begin{document}

\title{Unsupervised Variational Acoustic Clustering\\
\thanks{This work was supported by the Robust AI for SafE (radar) signal processing (RAISE) collaboration framework between Eindhoven University of Technology and NXP Semiconductors, including a Privaat-Publieke Samenwerkingen-toeslag (PPS) supplement from the Dutch Ministry of Economic Affairs and Climate Policy.}
}

\makeatletter
\newcommand{\linebreakand}{%
  \end{@IEEEauthorhalign}
  \hfill\mbox{}\par
  \mbox{}\hfill\begin{@IEEEauthorhalign}
}
\makeatother

\newlength{\extrawidth}
\settowidth{\extrawidth}{\textit{iiiiiiiiiiiiiii}}

\author{\IEEEauthorblockN{Luan Vinícius Fiorio}
\IEEEauthorblockA{\textit{Department of Electrical Engineering} \\
\textit{Eindhoven University of Technology}\\
Eindhoven, The Netherlands \\
l.v.fiorio@tue.nl}
\and
\IEEEauthorblockN{Bruno Defraene}
\IEEEauthorblockA{\hspace{\extrawidth}\textit{NXP Semiconductors}\hspace{\extrawidth}\\
Leuven, Belgium \\
bruno.defraene@nxp.com}
\and
\IEEEauthorblockN{Johan David}
\IEEEauthorblockA{\hspace{\extrawidth}\textit{NXP Semiconductors}\hspace{\extrawidth}\\
Leuven, Belgium \\
j.david@nxp.com}
\and
\IEEEauthorblockN{Frans Widdershoven}
\IEEEauthorblockA{\hspace{\extrawidth}\textit{NXP Semiconductors}\hspace{\extrawidth}\\
Eindhoven, The Netherlands \\
frans.widdershoven@nxp.com}
\and
\IEEEauthorblockN{Wim van Houtum}
\IEEEauthorblockA{\hspace{\extrawidth}\textit{NXP Semiconductors}\hspace{\extrawidth}\\
Eindhoven, The Netherlands \\
wim.van.houtum@nxp.com}
\and
\IEEEauthorblockN{Ronald M. Aarts}
\IEEEauthorblockA{\textit{Department of Electrical Engineering} \\
\textit{Eindhoven University of Technology}\\
Eindhoven, The Netherlands \\
R.M.Aarts@tue.nl}
}

\maketitle

\begin{abstract}
We propose an unsupervised variational acoustic clustering model for clustering audio data in the time-frequency domain. The model leverages variational inference, extended to an autoencoder framework, with a Gaussian mixture model as a prior for the latent space. Specifically designed for audio applications, we introduce a convolutional-recurrent variational autoencoder optimized for efficient time-frequency processing. Our experimental results considering a spoken digits dataset demonstrate a significant improvement in accuracy and clustering performance compared to traditional methods, showcasing the model’s enhanced ability to capture complex audio patterns. 
\end{abstract}


\section{Introduction}
Unsupervised clustering is crucial in audio applications \cite{hershey2016deep}, particularly for hardware-constrained devices like hearing aids \cite{larry2018acoustic}, where different processing is applied per detected acoustic scene \cite{park2020speech}. Traditional methods struggle to model the complex, high-dimensional nature of audio signals, resulting in suboptimal clustering \cite{foote1999overview}. Variational autoencoders (VAEs) are a promising tool for the task since they are capable of learning more efficient, low-dimensional representations of data \cite{kingma2014autoencoding}.

Variational autoencoders are frequently employed in unsupervised learning tasks due to their ability to learn from data without the need for labels or ground truth \cite{kingma2014autoencoding}. Naturally, the approach was modified towards clustering, where the prior distribution of latent variables, commonly a multivariate Gaussian distribution, was changed to a multivariate Gaussian mixture model, allowing for clustering behavior \cite{ugur2020variational, dilokthanakul2017deep, jiang2017variational}. Variational clustering was successfully applied to image applications, more specifically using the MNIST dataset \cite{deng2012mnist}.

In audio, particularly in speech processing, variational autoencoders have been used for speech enhancement applications \cite{huajian2021variational}, where latent representations for the clean speech are obtained based on noisy samples, such that the model can reconstruct clean data. Given the importance of temporal dependencies in audio processing, alternative architectures to the VAE such as the stochastic temporal convolutional network \cite{richter2020speech} have been proposed for more effective speech processing. Moreover, for the task of unsupersived clustering, to the best of our knowledge, its application to audio signals using generative models was not considered before.

In this work, we propose an unsupervised variational acoustic clustering (UVAC) model, building on \cite{ugur2020variational} and \cite{kingma2014autoencoding} towards the unsupervised clustering of audio data. We consider a convolutional-recurrent neural network (NN) model, and facilitate temporal processing by inputting the NN with a window of time frames \cite{fiorio2024spectral}. In comparison to traditional approaches, UVAC substantially enhances unsupervised accuracy, normalized mutual information and other clustering metrics.

\section{Variational Inference}

Consider a dataset $\mathbf{X} = \{\mathbf{x}^{(i)}\}^{N}_{i=1}$ with $N$ independent and identically distributed samples. From Bayes theorem, we can perform (statistical) inference by obtaining the posterior distribution
\begin{equation}
\label{eq:bayes}
    p_\theta(\mathbf{z}|\mathbf{x}^{(i)}) = \frac{p_\theta(\mathbf{x}^{(i)}|\mathbf{z}) p_\theta(\mathbf{z})}{p_\theta(\mathbf{x}^{(i)})},
\end{equation}
where $\mathbf{z}$ is a latent variable that meaningfully represents the underlying distribution of the data, and $\theta$ are model parameters. However, for most practical cases, $p_\theta(\mathbf{x}^{(i)})$ is unknown. As a work around, the probability distribution $p_\theta(\mathbf{x}^{(i)})$ can be marginalized as
\begin{equation}
\label{eq:marginalization}
    p_{\theta}(\mathbf{x}^{(i)}) = \sum_{\mathbf{z}} p_{\theta}(\mathbf{x}^{(i)}|\mathbf{z})p_{\theta}(\mathbf{z}), 
\end{equation}
being intractable as it requires summing over all possible values of $\mathbf{z}$ -- often with high order or complex relations. 

To make \eqref{eq:marginalization} and \eqref{eq:bayes} tractable, we introduce a \emph{variational} distribution with parameters $\phi$ to approximate the intractable posterior, $q_\phi(\mathbf{z}|\mathbf{x}^{(i)}) \approx p_\theta(\mathbf{z}|\mathbf{x}^{(i)})$, which we apply to \eqref{eq:marginalization},
\begin{equation}
\label{eq:expandedmarginalization}
    p_{\theta}(\mathbf{x}^{(i)}) = \sum_{\mathbf{z}} q_\phi(\mathbf{z}|\mathbf{x}^{(i)}) \frac{p_{\theta}(\mathbf{x}^{(i)}|\mathbf{z})p_{\theta}(\mathbf{z})}{q_\phi(\mathbf{z}|\mathbf{x}^{(i)})},
\end{equation}
and define $\sum_{\mathbf{z}} q_\phi(\mathbf{z}|\mathbf{x}^{(i)})$ as the expectation over $q_\phi(\mathbf{z}|\mathbf{x}^{(i)})$, represented by $\mathbb{E}_{q_\phi(\mathbf{z}|\mathbf{x}^{(i)})}[\cdot]$, allowing us to rewrite \eqref{eq:expandedmarginalization} as
\begin{equation}
\label{eq:expectedmarginalization}
    p_{\theta}(\mathbf{x}^{(i)}) = \mathbb{E}_{q_\phi(\mathbf{z}|\mathbf{x}^{(i)})} \left[ \frac{p_{\theta}(\mathbf{x}^{(i)}|\mathbf{z})p_{\theta}(\mathbf{z})} {q_\phi(\mathbf{z}|\mathbf{x}^{(i)})} \right]\!\!.
\end{equation}

Based on \eqref{eq:marginalization}, we know that the model's ability to represent data $\mathbf{x}$ can be measured by its log-likelihood, thus, its maximization becomes a cost function for obtaining the model parameters $\theta$, and is defined as follows:
\begin{equation}
\label{eq:loglikelihood}
    \max_{\theta} \log p_{\theta}(\mathbf{X}) = \max_{\theta} \sum_{i=1}^{N} \log p_{\theta} (\mathbf{x}^{(i)}).
\end{equation}
Nevertheless, we want to modify \eqref{eq:expectedmarginalization} in such a way that we are able to arrive at a similar expression as \eqref{eq:loglikelihood}, which can be used as a lower bound for optimization. Moreover, applying the $\log$ operator to  \eqref{eq:expectedmarginalization} gives
\begin{equation}
\label{eq:logexpectedmarginalization}
    \log p_{\theta}(\mathbf{x}^{(i)}) = \log \mathbb{E}_{q_\phi(\mathbf{z}|\mathbf{x}^{(i)})} \left[ \frac{p_{\theta}(\mathbf{x}^{(i)}|\mathbf{z})p_{\theta}(\mathbf{z})} {q_\phi(\mathbf{z}|\mathbf{x}^{(i)})} \right]\!\!,
\end{equation}
which allows us for using Jensen's inequality since $\log$ is a concave function, defining the \emph{variational lower bound} $\mathcal{L}$:
\begin{multline}
\label{eq:variationallowerbound}
    \log p_{\theta}(\mathbf{x}^{(i)}) \geq \mathbb{E}_{q_\phi(\mathbf{z}|\mathbf{x}^{(i)})} \left[ \log p_{\theta}(\mathbf{x}^{(i)}|\mathbf{z}) + \log \frac{p_{\theta}(\mathbf{z})}{q_\phi(\mathbf{z}|\mathbf{x}^{(i)})} \right] \\
    = \mathbb{E}_{q_\phi(\mathbf{z}|\mathbf{x}^{(i)})} \left[ \log p_{\theta}(\mathbf{x}^{(i)}|\mathbf{z}) \right] - D_{KL} ( q_\phi(\mathbf{z}|\mathbf{x}^{(i)}) \, || \, p_{\theta}(\mathbf{z})) \\
    = \mathcal{L}^{(i)}(\theta, \phi).
\end{multline}

In \eqref{eq:variationallowerbound}, $\mathbb{E}_{q_\phi(\mathbf{z}|\mathbf{x}^{(i)})} \left[ \log p_{\theta}(\mathbf{x}^{(i)}|\mathbf{z}) \right]$ represents the reconstruction error of $\mathbf{x}$ from $\mathbf{z}$, i.e., how well the latent variables $\mathbf{z}$ explain the data $\mathbf{x}$. The $D_{KL}$ term $D_{KL} ( q_\phi(\mathbf{z}|\mathbf{x}^{(i)}) \, || \, p_{\theta}(\mathbf{z}))$ stands for the Kullback-Leibler (KL) divergence, which quantifies the difference between the prior $p_{\theta}(\mathbf{z})$ and the variational distribution $q_\phi(\mathbf{z}|\mathbf{x}^{(i)})$. The objective of variational inference is to maximize $\mathcal{L}$ in terms of $\theta$ and $\phi$, corresponding to finding a good approximation $q_\phi(\mathbf{z}|\mathbf{x}^{(i)})$ for the posterior $p_\theta(\mathbf{z}|\mathbf{x}^{(i)})$.

\section{Unsupervised variational Clustering}

Given the complexity of the problem, we consider a variational autoencoder \cite{kingma2014autoencoding} framework. The VAE is composed by a neural network encoder, which takes the data $\mathbf{x}$ to a latent representation $\mathbf{z}$, and a NN decoder that generates data based on the latent variables. 

To allow clustering behavior, we choose a multivariate Gaussian mixture model (GMM) prior $p_\varphi(\mathbf{z})$, replacing $p_\theta(\mathbf{z})$ in \eqref{eq:variationallowerbound} for the latent space \cite{ugur2020variational}, given by
\begin{equation}
\label{eq:gmmprior}
    p_\varphi(\mathbf{z}) = \sum_{c} \pi_c \ \mathcal{N}(\mathbf{z}; \boldsymbol{\mu}_c, \boldsymbol{\Sigma}_c), 
\end{equation}
where $c$ is one component in the mixture -- a cluster. The multivariate GMM defined in \eqref{eq:gmmprior} is based on multivariate Gaussians $\mathcal{N}$ with means $\boldsymbol{\mu}_c$ and variances $\boldsymbol{\Sigma}_c$, whose dimension is the same as that of $\mathbf{z}$ ($d_\mathbf{z}$), linearly combined by a weighting vector $\pi_c$. The GMM parameters are condensed in a variable $\varphi \coloneqq \{\pi_c, \boldsymbol{\mu}_c, \boldsymbol{\Sigma}_c\}_{c=1}^{C}$, for $C$ components. 

\begin{figure}[!t]
\centering
    \begin{subfigure}{.45\textwidth}
        \centering
        \includegraphics[width=0.8\textwidth]{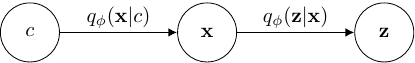}  
        \caption{Inference model}
        \label{fig:inferencemodel}
        \vspace{5mm}
    \end{subfigure}
    \begin{subfigure}{.45\textwidth}
        \centering
        \includegraphics[width=0.8\textwidth]{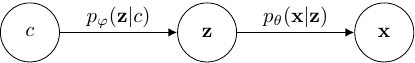}
        \caption{Generative model}
        \label{fig:generativemodel}
    \end{subfigure}
\caption{Inference and generative models for unsupervised clustering.}
\label{fig:models}
\vspace{-5mm}
\end{figure}

The considered inference and generative models are presented in Fig.~\ref{fig:models}, where we consider two assumptions: i) for inference, the data are originated from component $c$ of an unknown GMM $q_\phi(\mathbf{x}^{(i)}|c)$; and ii) the latent variables have a GMM prior $p_\varphi(\mathbf{z})$, as described in \eqref{eq:gmmprior}, for the generation of data. The inference model shown in Fig.~\ref{fig:inferencemodel} is given by the encoder $q_\phi(\mathbf{z}|\mathbf{x}^{(i)}) = \mathcal{N}(\mathbf{z}; \boldsymbol{\mu}_\phi, \boldsymbol{\Sigma}_\phi)$, where a neural network $f$ with parameters $\phi$ and input $\mathbf{x}^{(i)}$ generates the multivariate output $f_\phi(\mathbf{x}^{(i)}) = [\boldsymbol{\mu}_\phi, \boldsymbol{\Sigma}_\phi]$. On the other hand, the generative model from Fig.~\ref{fig:generativemodel} is composed by a NN decoder $g$ with parameters $\theta$ that takes $\mathbf{z}$ as input and is described as $p_\theta(\mathbf{x}^{(i)}|\mathbf{z}) = [\hat{\mathbf{x}}^{(i)}] = g_\theta(\mathbf{z})$, with $\hat{\mathbf{x}}^{(i)}$ being the generated data.

\begin{figure*}[!ht]
    \centering
    \includegraphics[width=0.825\textwidth]{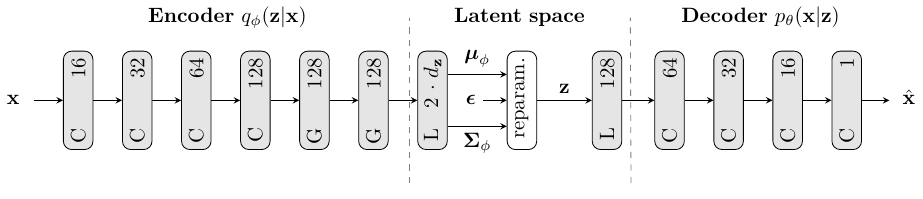}
    \vspace{-4mm}
    \caption{Schematic of the proposed convolutional-recurrent variational autoencoder. The number in each layer indicates output channels. The $\mathrm{C}$ encoder layers consist of Conv2D with BatchNorm2D and ReLU functions in all layers. The $\mathrm{G}$ layers are gate recurrent units (GRUs). The layers of the latent space are linear ($\mathrm{L}$) without activation. The decoder layers are Conv2D.T, with ReLU activation in all layers but the last, with sigmoid. All $\mathrm{C}$ kernels are (8,8) with stride (2,2) and padding (3,3).}
    \label{fig:VAE}
    \vspace{-5mm}
\end{figure*}

All parameters of the model ($\theta$, $\phi$, and $\varphi$) are optimized simultaneously to maximize the variational lower bound, modified from \eqref{eq:variationallowerbound} to meet the GMM prior:
\begin{multline}
\label{eq:variationallowerboundclustering}
    \mathcal{L}^{(i)}(\theta, \phi, \varphi) = \\
    \mathbb{E}_{q_\phi(\mathbf{z}|\mathbf{x}^{(i)})} \left[ \log p_{\theta}(\mathbf{x}^{(i)}|\mathbf{z}) \right] - D_{KL} ( q_\phi(\mathbf{z}|\mathbf{x}^{(i)}) \, || \, p_\varphi(\mathbf{z})).
\end{multline}
In practice, to allow gradient flow, the reconstruction error in \eqref{eq:variationallowerboundclustering} can be obtained with the reparametrization trick \cite{kingma2014autoencoding} and Monte Carlo sampling \cite{ugur2020variational}:
\begin{equation}
\label{eq:reconstructionloss}
    \mathbb{E}_{q_\phi(\mathbf{z}|\mathbf{x}^{(i)})} \left[ \log p_{\theta}(\mathbf{x}^{(i)}|\mathbf{z}) \right] \approx \frac{1}{M} \sum_{m} p_{\theta}(\mathbf{x}^{(i)}|\mathbf{z}_m),
\end{equation}
with $m$ being a Monte Carlo sample out of $M$, and $\mathbf{z}_m = \boldsymbol{\mu}_\phi + \boldsymbol{\Sigma}_{\phi}^{1/2} \boldsymbol{\epsilon}_m$ (reparametrization trick), with an auxiliary random variable $\boldsymbol{\epsilon}_m \sim \mathcal{N}(\mathbf{0}, \mathbf{I})$. The KL divergence term of \eqref{eq:variationallowerboundclustering}, i.e., the divergence between a single component multivariate Gaussian and a multivariate GMM with $C$ components, cannot be calculated in closed-form, but it can be approximated \cite{ugur2020variational, hershey2007approximating}. Thus, we approximate \eqref{eq:variationallowerboundclustering} considering that $\boldsymbol{\Sigma}_\phi = \text{diag}(\{\Sigma_{\phi, j}\}_{j=1}^{d_\mathbf{z}})$ and $\boldsymbol{\Sigma}_c = \text{diag}(\{\Sigma_{c, j}\}_{j=1}^{d_\mathbf{z}})$ as
\begin{multline}
    D_{KL} ( q_\phi(\mathbf{z}|\mathbf{x}^{(i)}) \, || \, p_\varphi(\mathbf{z})) \approx \\
    -\log \sum_{c} \pi_c \exp \Biggl( -\frac{1}{2} \sum_{j} \Biggl[ \frac{(\mu_{\phi,j} - \mu_{c,j})^2}{\Sigma_{c,j}} \\ 
    + \log \frac{\Sigma_{c,j}}{\Sigma_{\phi,j}} 
    - 1 + \frac{\Sigma_{\phi,j}}{\Sigma_{c,j}} \Biggr] \Biggr)\!.
\end{multline}

Moreover, we define the cost function to be optimized as the average of \eqref{eq:variationallowerboundclustering} over all samples of the dataset $\mathbf{X}$, as
\begin{equation}
\label{eq:costfunction}
    \mathcal{J}(\theta, \phi, \varphi) = \frac{1}{N} \sum_{i=1}^{N} \mathcal{L}^{(i)}(\theta, \phi, \varphi),
\end{equation}
resulting in the optimization problem
\begin{equation}
\label{eq:optimizationproblem}
    \max_{\theta, \phi, \varphi} \mathcal{J}(\theta, \phi, \varphi) \Leftrightarrow \min_{\theta, \phi, \varphi} -\mathcal{J}(\theta, \phi, \varphi),
\end{equation}
which right-hand side can be minimized via backpropagation.

\section{Architecture}

Unsupervised clustering using a variational autoencoder framework was previously applied to image applications \cite{jiang2017variational, dilokthanakul2017deep, ugur2020variational}, specifically for the classification of handwritten digits with the MNIST dataset \cite{deng2012mnist}. For such a case, where the input dimension is relatively small, the authors have employed, successfully, fully connected NNs as encoder and decoder.

In our case, we consider the unsupervised clustering of audio signals, validating the proposal through the AudioMNIST dataset \cite{becker2023audiomnist}. Nevertheless, the input size is unwieldy for a fully connected NN, as it becomes inefficient given the unnecessary number of parameters. Envisioning a more efficient implementation, we define the autoencoder as a convolutional-recurrent network. Additionally, as audio strongly depends on time-correlations, we employ an explicit time-context at the input \cite{fiorio2024spectral} to enhance temporal processing. 

\subsection{Data}
\label{ssec:data}
We consider the problem of unsupervised clustering of spoken digits with the AudioMNIST dataset \cite{becker2023audiomnist}. The dataset contains 30000 audio samples -- of which 24000 are randomly selected for training, 3000 for validation, and 3000 for testing -- where each file contains the audio recording, at 16 kHz, of a spoken digit. The speakers are of different gender and age. The raw audio data are pre-processeded as follows. First, we pad zeros to each audio sample until the desired duration of 1 second is achieved. The padded audio is then applied to a Short-term Fourier Transform (STFT), with length of 960 samples, Hann window of the same size, and a hop of 480 samples. Moreover, we take the module of the output of the STFT and limit the frequency range to 128 frequency bins -- approximately 6 kHz. Such a frequency range has showed to be sufficient for classification tasks on AudioMNIST in preliminary tests. Finally, the spectrograms are normalized by their mean and variance, with ranges limited from 0 to 1 by a min-max adjustment.

The input to the neural network is the padded and pre-processeded full second of audio, containing 128 frequency bins and 99 time bins. The zero-frequency bin is removed from all samples. By feeding the network with the entire duration of a file, we leverage time dependencies, as observed in \cite{fiorio2024spectral}. Importantly, we generate an activity detection vector containing zeros in the zero-padded time-indexes, and ones otherwise. This vector multiplies both target and prediction during the calculation of the reconstruction loss. If this is not done, the model might learn to classify zero-padding duration.

\subsection{Autoencoder}
\label{ssec:autoencoder}

Generative models tend to require a massive number of parameters to achieve desirable performance \cite{kaplan2020scalinglawsneurallanguage}. To form a model with a reduced number of parameters, we define a convolutional-recurrent variational autoencoder composed of a convolutional-recurrent encoder $q_\phi (\mathbf{z}|\mathbf{x}^{(i)})$, a latent space with prior $p_\varphi(\mathbf{z})$, and a convolutional decoder $p_\theta(\mathbf{x}^{(i)}|\mathbf{z})$, as shown in Fig.~\ref{fig:VAE}. We refer to the proposed model, for simplicity, as the unsupervised variational acoustic clustering (UVAC) model. Notice that when defining the architecture, recurrent layers were necessary as removing them would make the performance insufficient in terms of accuracy.

We set the latent space dimension as $d_\mathbf{z}=10$ because early experiments indicated that this value is effective for unsupervised clustering on the AudioMNIST dataset. Nevertheless,     linear layers are used in the latent space to convert the number of channels to the chosen latent dimension, a necessary modification in the proposed model when compared to fully connected VAEs in literature \cite{ugur2020variational}. The first linear layer converts the number of channels from 128, at the end of the encoder, to $2 \cdot d_\mathbf{z}$, where the first $d_\mathbf{z}$ data points are taken as values for the (multivariate) mean, and the last $d_\mathbf{z}$ for the variance. The second linear layer converts the latent dimension to 128 channels. Moreover, as previously described, the prior distribution for the latent space $p_\varphi(\mathbf{z})$ is a multivariate Gaussian mixture model. For our experiments, we choose $C=10$ GMM components, which is the same as the number of classes in the dataset.

We initialize all convolutional, recurrent, and linear layers with the Kaiming uniform initialization for the weights and zeros for the biases. The GMM prior $p_\varphi(\mathbf{z})$, defined in \eqref{eq:gmmprior}, is initialized as follows. The weighting vector $\pi_c$ takes a uniform distribution, with lower and upper bound 0.0 and 1.0, respectively. The vector of means $\boldsymbol{\mu}_c$ is initialized with the Xavier uniform initialization, and the variances $\boldsymbol{\Sigma}_c$ are all set as zero. Importantly, the clusters are chosen based on the probability of data point $\mathbf{x}^{(i)}$ belonging to the $c$th cluster \cite{ugur2020variational}:
\begin{multline}
    q_{\phi}(c|\mathbf{x}^{(i)}) = p_\theta(c|\mathbf{z}) = \frac{p_\varphi(c) p_\varphi(\mathbf{z}|c)}{p_\varphi(\mathbf{z})} = \\
    \frac{\pi_c \ \mathcal{N}(\mathbf{z}; \boldsymbol{\mu}_c, \boldsymbol{\Sigma}_c)}{\sum_c \pi_c \ \mathcal{N}(\mathbf{z}; \boldsymbol{\mu}_c, \boldsymbol{\Sigma}_c)}.
\end{multline}

During training, the complete model is used. For applying the NN model in inference mode, however, only the encoder and the latent space are necessary, because we perform clustering based on the GMM prior. Therefore, the inference UVAC model has much less parameters than the version used in training. In numbers, the complete model has 2M parameters, while roughly 1.3M are used for inference. Although this number of parameters is low when compared to other generative approaches \cite{kaplan2020scalinglawsneurallanguage}, further studies in model size reduction are of interest, but out of scope for this paper.

\section{Experimental evaluation}

In this section we present the results obtained with the UVAC model described in Section~\ref{ssec:autoencoder}. We trained the model for 500 epochs using the Adam optimizer, with an initial learning rate of 0.005, exponentially decaying until the final value of 0.0005 at the last epoch. The batch size considered for training was of 64 data points. Differently from \cite{ugur2020variational}, we keep both terms in \eqref{eq:variationallowerboundclustering} with equal weight. During tests, we noticed that the performance is very sensible to the weighting of the KL divergence term, and for the AudioMNIST dataset, a weighting of 1.0 for each term achieved the best results. Additionally, we keep $M=1$ in \eqref{eq:reconstructionloss} for efficiency.

\subsection{Metrics}
In the following we describe the considered metrics for unsupervised clustering.

\subsubsection{Unsupervised accuracy}
in unsupervised clustering tasks, the numeric labels may not correspond directly to the ground truth labels. We then consider an unsupervised approach for calculating accuracy, which consists of finding the matching truth labels for the clusters via the Hungarian algorithm \cite{kuhn1955hungarian}. Unsupervised accuracy ranges from 0 to 100\%.

\subsubsection{Normalized mutual information}
the normalized mutual information (NMI) is an information theoretic approach that evaluates the clustering quality by measuring the amount of shared information between clustering assignments and truth labels \cite{vinh2010information}. Its range is from 0 to 1.

\subsubsection{Silhouette score}
the Silhouette score \cite{rousseeuw1987silhouettes} measures how similar a data point is to its own cluster in comparison to other clusters. It combines cohesion (how close data points within a cluster are) and separation (how distinct is a cluster from another). The range is from -1 to +1: -1 indicates misclassification; 0 tells us that clusters overlap; and +1 indicate optimal clustering.

\subsubsection{Davies-Bouldin index}
the Davies-Boudlin Index (DBI) \cite{davies1979cluster} is defined as the average similarity ratio of each cluster with the cluster that is most similar to it. A lower DBI indicates better clustering -- in terms of compactness and separation. Its range can vary from 0 to infinity.


\subsection{Other methods}

For the sake of comparison, we perform unsupervised clustering using two traditional methods, named K-means \cite{hartigan1979kmeans} and the optimization of a Gaussian mixture model using the expectation-maximization (EM) algorithm (GMM-EM) \cite{dempster1977em}. Other classical approaches and derivations are assumed to achieve similar performance as K-means and GMM-EM. We also tried to apply dimensionality reduction techniques on the pre-processed data with the traditional methods, but no performance improvement was obtained. 

To the best of our knowledge, this is the first study to apply variational autoencoders for the unsupervised clustering of audio data. While other generative models are capable of clustering, they have not been specifically employed for audio data clustering as we tackle in this paper.

\begin{table}[!t]
    \centering
    \caption{Unsupervised clustering metrics on the test set of AudioMNIST, either considering labels as clusters or by applying K-means, GMM optimized by EM, and UVAC as clustering methods, averaged over 10 independent runs.}    
    \begin{tabular}{l c c c c}
    \hline
        \textbf{Method} & \textbf{Accuracy} (\%) & \textbf{NMI} & \textbf{Silhouette} & \textbf{DBI} \\
    \hline        
        \textcolor{gray}{None (labels)} & \textcolor{gray}{100.00} & \textcolor{gray}{1.00} & \textcolor{gray}{-0.04} & \textcolor{gray}{5.56} \\     
        K-means & 18.40 & 0.10 & 0.13 & 2.04 \\ 
        GMM-EM & 17.62 & 0.09 & 0.13 & 1.95 \\
        UVAC & \textcolor{tueScharlaken}{70.78} & \textcolor{tueScharlaken}{0.71} &\textcolor{tueScharlaken}{0.21} & \textcolor{tueScharlaken}{1.61} \\            
    \hline
    \end{tabular}
    \label{tab:results}
    \vspace{-5mm}
\end{table}

\begin{figure*}[!t]
\centering
    \begin{subfigure}{.3\textwidth}
        \centering
        \includegraphics[width=0.99\textwidth]{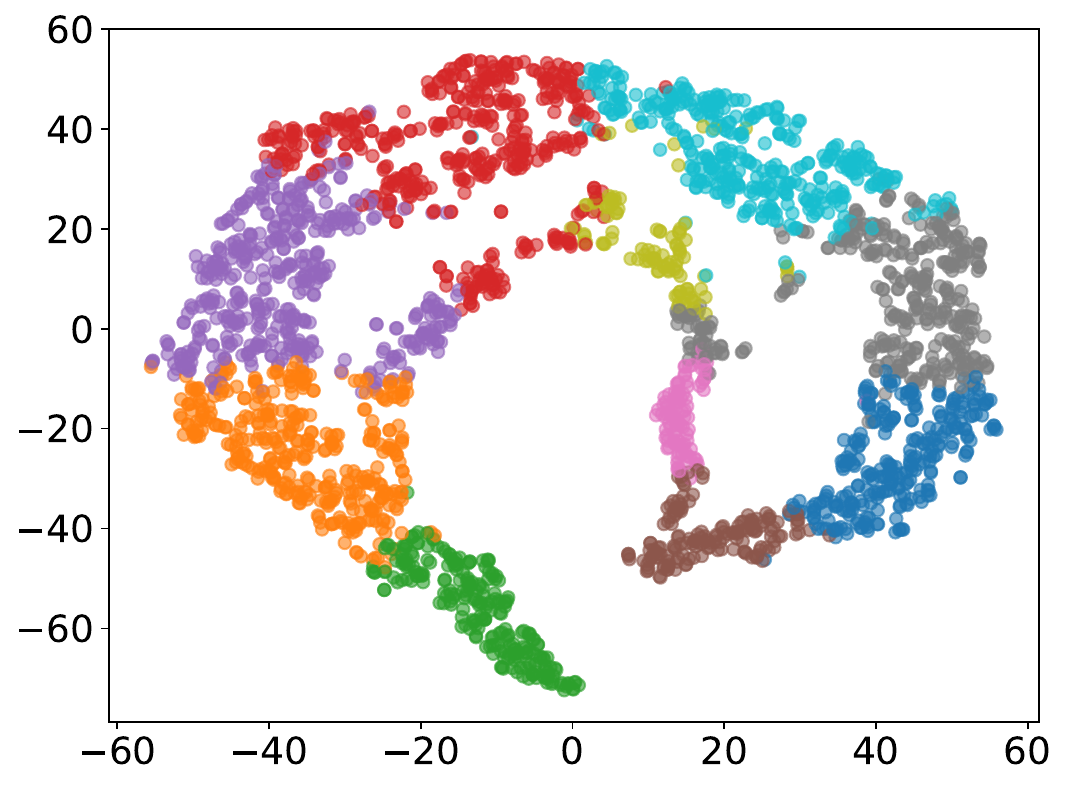}  
        \caption{K-means}
        \label{fig:kmeans_clusters}
    \end{subfigure}
    \begin{subfigure}{.3\textwidth}
        \centering
        \includegraphics[width=0.99\textwidth]{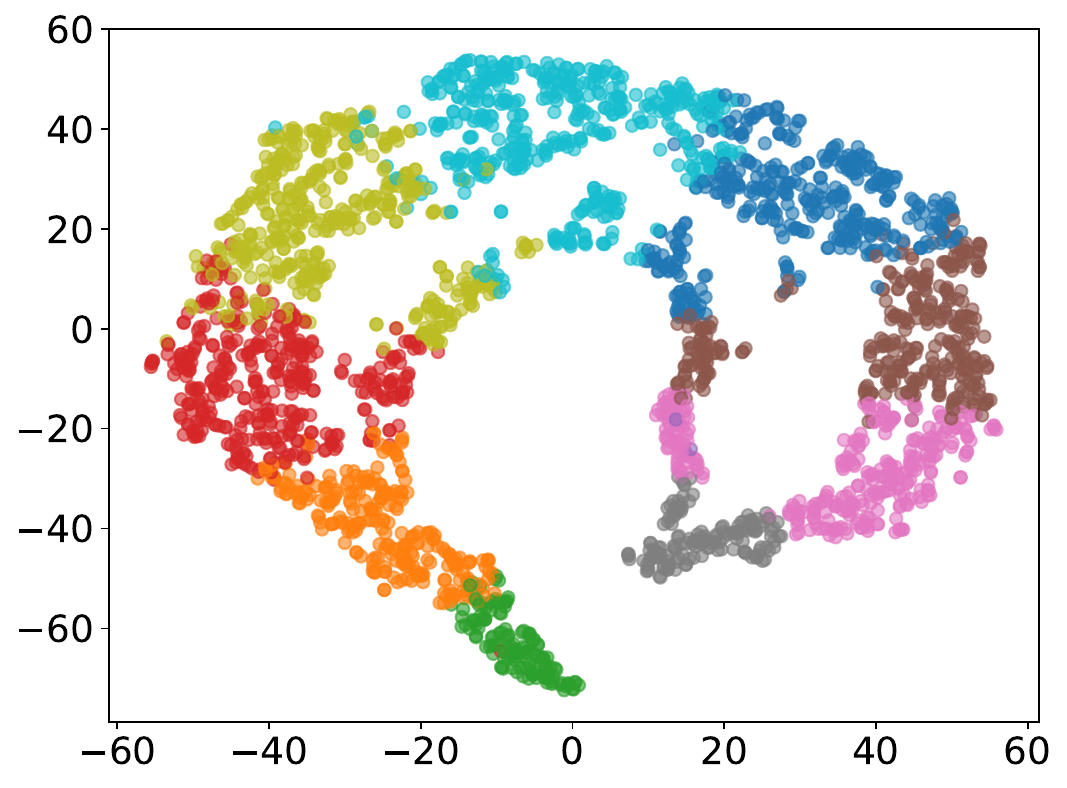}
        \caption{GMM-EM}
        \label{fig:gmm_clusters}
    \end{subfigure} 
    \begin{subfigure}{.3333\textwidth}
        \centering
        \includegraphics[width=0.99\textwidth]{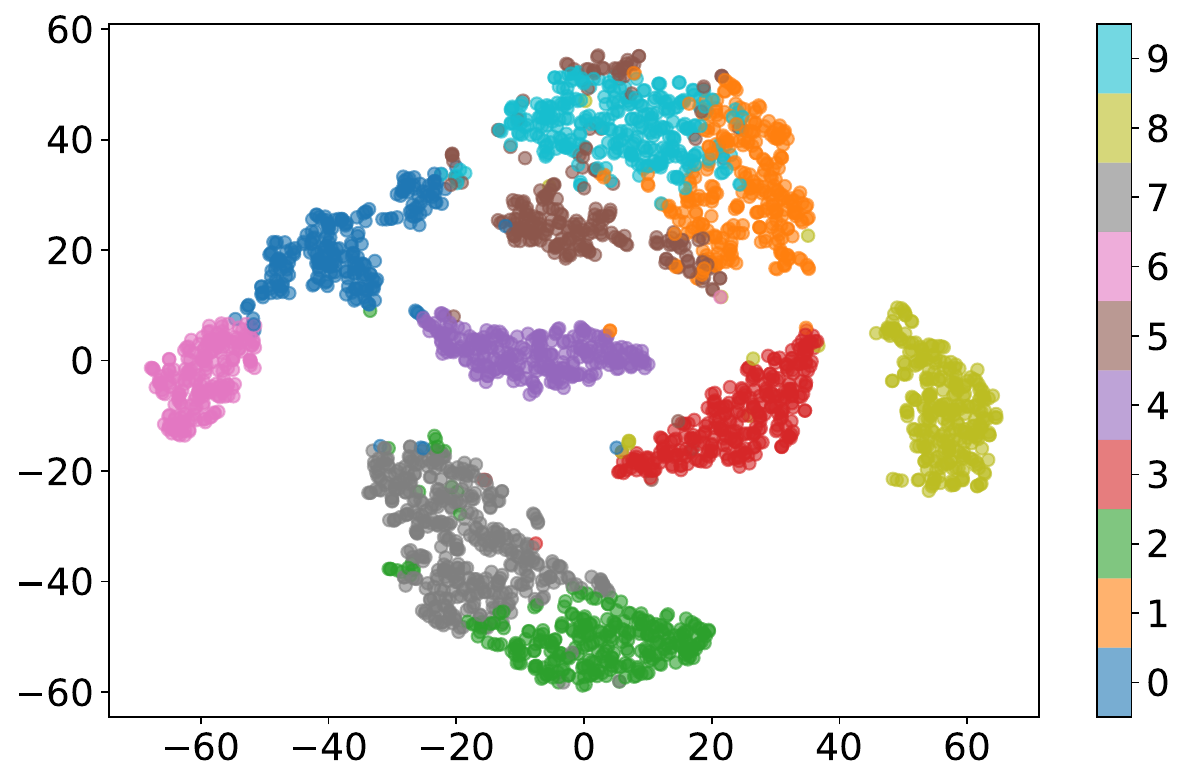}
        \caption{UVAC}
        \label{fig:uvac_clusters}
    \end{subfigure}
\caption{Clusters obtained for the AudioMNIST test set. The data size is reduced for plotting using t-distributed stochastic neighbor embedding. Each color represents a cluster (labels shown in bar plot), and each circle in the plot is a data point.}
\label{fig:clusters}
\vspace{-5mm}
\end{figure*}

\subsection{Results}

The obtained results are shown in Table~\ref{tab:results}. For the labels, K-means, and GMM-EM, we calculate the metrics with the STFT data, while for UVAC we use the latent variable $\mathbf{z}$. First, notice how the classes provided with the AudioMNIST dataset, referent to the spoken digits, do not form good clusters in terms of the Silhouette score and DBI. Improvements are clear when K-means and GMM-EM are applied, raising the Silhouette score and lowering the DBI, but dramatically reducing unsupervised accuracy and normalized mutual information. Such methods, therefore, show themselves insufficient for an accurate unsupervised clustering when data are of high dimensionality or contain complex relations. On the other hand, when UVAC is applied, accuracy and NMI are increased to approximately 71\%, and the unsupervised clustering metrics are enhanced if compared to the other approaches. 

Based on these results, for audio applications in general, we expect the accuracy to be limited when unsupervised clustering objectives are involved, almost as an abstract form of regularization. A perfect match with the truth labels does not provide good clusters, as we can see from the metrics obtained with the labels. Furthermore, the sufficiently high accuracy (around 71\%) obtained by the proposed model shows that the digit being spoken, alone, is a major part of the data features, but accompanied by other features present in the STFTs, e.g., voice frequency, microphone noise, time to pronounce a number, etc. This ``regularizing'' effect is mainly caused by the complexity of the audio data, which is not observed for image processing \cite{ugur2020variational, dilokthanakul2017deep, jiang2017variational}. This also makes the inclusion of annealing strategies in the loss function nontrivial.

Additionally, in Fig.~\ref{fig:clusters} we show an example of the clusters obtained for AudioMNIST's test set. Visually, K-means and GMM-EM clusters are similar, being insufficient in terms of compactness and separation. On the other hand, UVAC's clusters are more spread and compact for most classes. Clearly, the traditional methods are limited in their ability to capture the (complex) underlying data structure. The incorporation of advanced clustering strategies, like UVAC, when dealing with audio data, might drastically improve the performance of a system that depends on unsupervised clustering.

\section{Conclusion}

We proposed a variational autoencoder model for unsupervised acoustic clustering of audio data. The proposed model is a convolutional-recurrent variational autoencoder with linear layers in the latent space, which has a multivariate GMM prior. The obtained clustering results with the UVAC model over the AudioMNIST dataset show substantial improvements in relation to other approaches. The UVAC is capable of maintaining high accuracy and normalized mutual information while increasing clustering quality. For future works, we suggest the development of a robust annealing strategy of the loss terms aimed at audio applications.

\bibliographystyle{IEEEtran}
\bibliography{references}

\end{document}